\documentclass{webofc}

\usepackage[varg]{txfonts}   
\usepackage{hyperref}

\usepackage[utf8]{inputenc}
\hypersetup{colorlinks=true,citecolor=blue,urlcolor=blue,linkcolor=blue}
\begin{document}
\title{Initial spin fluctuations in heavy-ion collisions and where to find them}
\author{\firstname{Giuliano} \lastname{Giacalone}\inst{1} \and
\firstname{Enrico} \lastname{Speranza}\inst{2}
}

\institute{Theoretical Physics Department, CERN, 1211 Geneva 23, Switzerland 
\and
Department of Physics and Astronomy, University of Florence, Via G. Sansone 1, 50019\\Sesto Fiorentino, Italy}
\abstract{Collective spin phenomena in the final states of heavy-ion collisions are typically understood to originate from vorticity and shear in the quark-gluon plasma (QGP). Here, we ask whether spin could already be present in the initial condition of the collisions. In particular, we argue that if a spin density exists at the beginning of the QGP expansion, it should experience event-by-event fluctuations due to the finite number of participant nucleons. In this contribution, we propose a simple model of fluctuating spin initial conditions for event-by-event spin hydrodynamics based on the Glauber Monte Carlo paradigm. We postulate that, if the net spin of the events is conserved from the initial to the final state, then initial state fluctuations of spin should manifest in specific spin correlations of $\Lambda$ hyperons. Within our picture, we predict that this signal is much larger in central O+O collisions than in central Pb+Pb collisions.}
\maketitle
\section{Introduction}
\label{intro}

The observation of hadron spin polarization in heavy-ion collisions has highlighted the importance of spin degrees of freedom in the dynamics of hot and dense QCD matter, opening a new direction for understanding collective spin phenomena in relativistic quantum fluids \cite{STAR:2017ckg,ALICE:2019onw,ALICE:2019aid,STAR:2022fan,HADES:2022enx,CMS:2025nqr,Becattini:2024uha,Niida:2024ntm}. Most theoretical descriptions assume local equilibrium, where polarization arises solely from thermal vorticity or shear at freeze-out \cite{Becattini:2024uha}. This framework successfully predicts the global polarization of $\Lambda$ hyperons, whereas the description of the local polarization is more limited and continues to pose open questions. However, another fundamental challenge remains: the role of spin in the initial conditions of the collisions is not yet explored.

In this work~\cite{Giacalone:2025bgm}, we postulate the existence of an initial spin density and discuss the main features we expect it to present. Our point is that it is essential to account for fluctuations in the number of colliding nucleons, in agreement with the paradigm of initial-state fluctuations in heavy-ion collisions \cite{Alver:2010gr,Giacalone:2023hwk}. We provide a simple model for the fluctuating spin density, and discuss how the imprints of initial-spin fluctuations could be detected experimentally.


\begin{figure}[t]
    \centering
    \includegraphics[width=.9\linewidth]{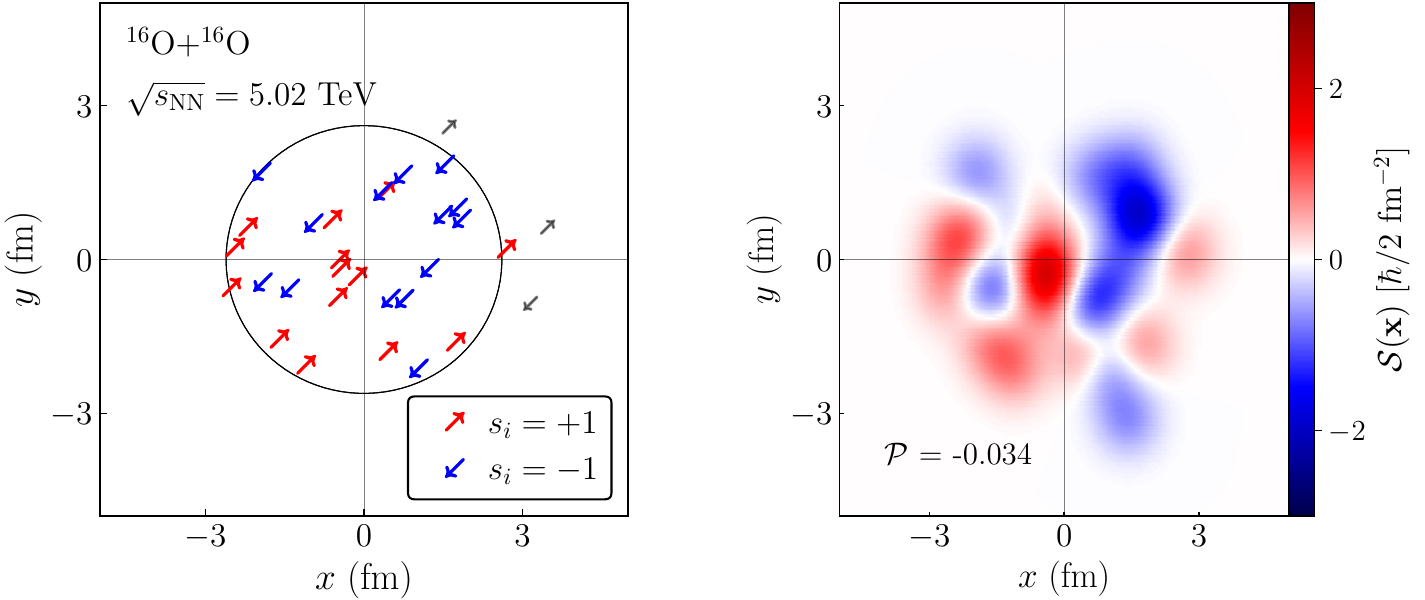}
    \caption{\textit{Left:} Transverse view of an O+O collision at $\sqrt{s_{\rm NN}}=5$~TeV. Each nucleon in the colliding ions is assigned a random spin projection, either up or down. Since $^{16}$O has $J=0$ in the ground state, it contains $A/2$ spin-up and $A/2$ spin-down nucleons. The spins are aligned along a random quantization axis. Colored arrows represent participant nucleons. \textit{Right:} The corresponding spin density, $\mathcal{S}({\bf x})$, as defined in Eq.~(\ref{eq:Sdens}). Here, the spin profile of each participant, $w_s({\bf x})$, is modeled as a two-dimensional Gaussian with width 0.5 fm, and $S_0=1$. The event shown yields $\int_{\bf x} \mathcal{S}({\bf x})=- \,\hbar/2$ and $\mathcal{P}=-0.034$.}
    \label{fig:1}
\end{figure}

\section{Model of spin density at midrapidity}

To construct the spin density, we follow a Glauber-type picture in which the scattering of two nuclei acts as a quantum measurement of their ground-state wavefunctions, characterized by position, spin, and isospin degrees of freedom. While the impact of fluctuations in nucleon positions (and isospins) in the initial state of heavy-ion collisions is well established, spin has not yet been explored in this context. We therefore adopt a scenario in which each scattering nucleon carries a definite spin projection, either up or down, with a common quantization axis for both nuclei. This leads to the configuration shown in the left panel of Fig.~\ref{fig:1} for a central O+O collision, where the colored arrows are the participant nucleons of the event.

From this assignment, we construct the spin density at midrapidity:
\begin{equation}
\label{eq:Sdens}
\mathcal S ({\bf x}) \equiv S_0 \frac{\hbar}{2} \sum_{i=1}^{N_{\rm part}} s_i \, w_s({\bf x}-{\bf x}_i) \,, \hspace{40pt} \mathcal{S} = \int d^2{\bf x}\,\mathcal{S}({\bf x}) = S_0 \frac{\hbar}{2} \sum_{i=1}^{N_{\rm part}} s_i \,,
\end{equation}
where ${\bf x}$ is a transverse coordinate, $w_s({\bf x}-{\bf x}_i)$ is the two-dimensional profile of spin around participant $i$, $s_i$ is the spin projection of that participant ($s_i = \pm 1$), $S_0$ is a dimensionless parameter controlling the fraction of nucleon spin deposited at midrapidity, while $\mathcal{S}$ is the net spin of the event. For the event shown in Fig.\ref{fig:1}, the corresponding $\mathcal S({\bf x})$ is shown in the right panel. This spin profile naturally leads to an event-wise polarization per participant, which is the central object of our analysis. This is defined as:
\begin{equation}
\label{eq:P}
\mathcal{P} (S_0 = 1) = \frac{1}{\hbar/2}\frac{\mathcal{S}}{N_{\rm part}}, \hspace{30pt} -1 \leq \mathcal{P} \leq 1.
\end{equation}
Averaging over many events yields $\langle \mathcal P \rangle = 0$, but the event-by-event fluctuations of $N_{\rm part}$ generate a non-zero standard deviation. To quantify this, we simulate minimum-bias Pb+Pb, Xe+Xe, and O+O collisions and we evaluate $\mathcal{P}$ on an event-by-event basis. We then compute ${\rm std}(\mathcal{P}) = \sqrt{\langle \mathcal P^2 \rangle}$
as a function of the centrality (defined from the initial entropy of the events). 

\begin{figure*}
    \centering
    \includegraphics[width=.9\linewidth]{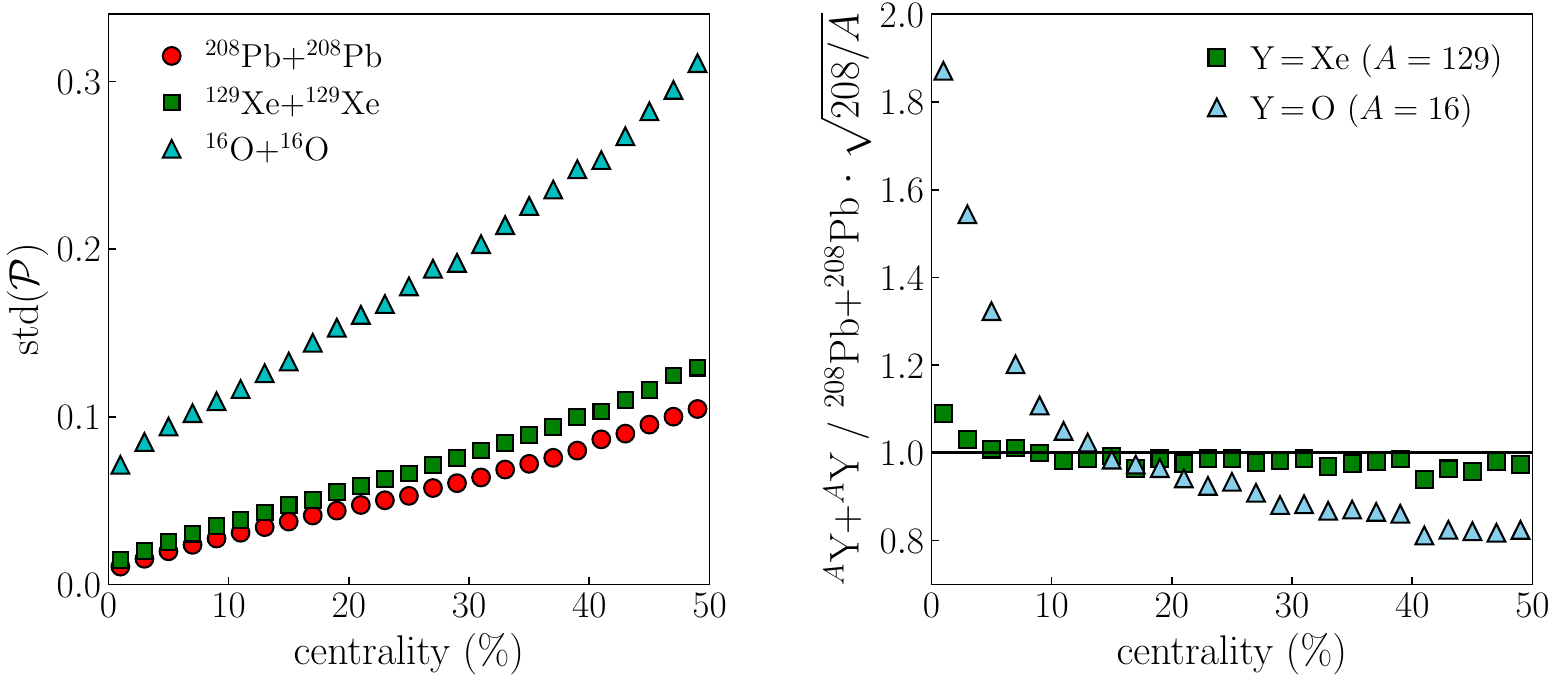}
    \caption{\textit{Left:} Standard deviation of the polarization parameter, $\mathcal{P}$, as defined by Eq.~(\ref{eq:P}) with $S_0=1$, in PB+Pb collisions (circles) Xe+Xe collisions (squares), and O+O collisions (Triangles) as a function of the collision centrality. \textit{Right:} Xe+Xe (squares) and O+O (triangles) results normalized by the Pb+Pb result, and rescaled by a factor $\sqrt{208/A}$, where $A$ is either 129 or 16.}
    \label{fig:2}
\end{figure*}

The result is in Fig.~\ref{fig:2} (for $S_0=1$), and highlights important qualitative features of this quantity. First, on the left-hand panel we see a clear $1/\sqrt{N_{\rm part}}$ scaling, increasing by about an order of magnitude from central to peripheral collisions, and similarly as one move from large Pb+Pb to small O+O collision systems. On the right-hand panel, we study the scaling with system size in more detail. Remarkably, $1/\sqrt{N_{\rm part}}$ scaling is nearly perfect as we move from Pb+Pb to Xe+Xe collisions, while it is strongly broken by O+O collisions.

\section{Observable spin correlation}

How is this analysis of relevance for experiments? Our point is that, if the net spin is conserved during the expansion of the QGP \cite{Florkowski:2017ruc,Speranza:2020ilk,Abboud:2025qtg}, the polarization per particle, $\mathcal{P}$, driven by the event-to-event net-spin fluctuations, should be carried over to the final states. The standard deviation $\sqrt{\langle \mathcal{P}^2 \rangle}$ can be measured through an appropriate two-particle spin correlation.

We start from the generic angular distribution of a proton emitted in a $\Lambda$ decay, $\frac{dN}{d\theta} \propto 1 + \alpha \,\mathcal{P}\,\cos\theta$, where $\theta$ is the angle between the proton momentum and the chosen quantization axis, $\alpha$ is the weak-decay parameter, while $\mathcal{P}$ is the polarization of the hyperon. Consider now a pair of $\Lambda$, both having the same polarization $\mathcal{P}$, and emitting protons independently (in their respective rest frames) with momenta $\vec{k}_1$ and $\vec{k}_2$, respectively. Denoting $\cos(\Delta\theta)=\vec{k}_1 \cdot\vec{k}_2$, the pair distribution differential in the relative angle can be written as:
\begin{equation}
\label{DD}
\frac{d^2N}{d\cos\theta_1 d\cos\theta_2} ~~ \longrightarrow~~\frac{dN}{d\cos(\Delta\theta)} \propto 1 + D \cos(\Delta\theta), \hspace{20pt} D = 3\langle \cos(\Delta\theta) \rangle \, ,
\end{equation}
where $D$ does not depend on the specific choice of Cartesian axes defining the quantization axis in the $\Lambda$ rest frame, and has been measured in high-energy physics analyses to probe quantum entanglement in top–antitop pairs \cite{ATLAS:2023fsd,CMS:2024pts}.

The average leading to the value of $D$ is for all $\Lambda$ hyperons measured in an individual event. Averaging over events, we extract an event-averaged spin–correlation coefficient,
\begin{align}
D = \alpha_1\alpha_2\mathcal{P}^2 ~~ \longrightarrow ~~ v_\Lambda^2 \equiv \frac{9}{\alpha_1 \alpha_2} \langle\!\langle \cos (\Delta \theta) \rangle\!\rangle = \frac{3}{\alpha_1 \alpha_2} \langle D \rangle = \langle \mathcal{P}^2 \rangle ,
\end{align}
where double brackets denote an average over all pairs in an event followed by an average over all events. This observable isolates the event-by-event fluctuation of the magnitude of the polarization, $\mathcal{P}$. Note that $v_\Lambda^2$ will also capture final-state effects related to the fluctuations of the vorticity field. However, at the highest collision energies, the latter effects are very small \cite{Pang:2016igs}, while even for $S_0\approx0.1$, we would still generate percent-level signals.

In addition, it would be crucial to establish whether the spin–spin correlation is a long-range effect. Existing studies of spin correlations typically focus on local phenomena at the freeze-out surface. However, if the initial spin density introduced here is approximately boost-invariant, the resulting correlations should persist even when the correlated $\Lambda$ hyperons are separated by large rapidities. Observing such a signal would provide conclusive evidence that the underlying mechanism originates in the early stages of the collisions.

 Finally, the proposed observable would probe an intriguing new object: the nuclear two-body density in spin space, formally defined as $\rho^{(2)}(s_1,s_2) = \sum_{s_3 \ldots s_A} \int_{{\bf x}_1, \ldots, {\bf x}_A} |\Psi({\bf x}_1, \ldots, {\bf x}_A, s_1, \ldots, s_A)|^2 \,$, where $\Psi$ is the nuclear wavefunction. In our model, nucleon spins are sampled independently, reducing this quantity to a product of one-body densities without spin-spin correlations. More sophisticated \textit{ab initio} approaches, such as Nuclear Lattice or Quantum Monte Carlo methods would however predict non-trivial spin structures. This could ultimately reveal how nucleon spin correlations imprint themselves on $\Lambda$ hyperon correlations. We leave this question for future work.

\section{Conclusions and outlook}

We have introduced a model for the initial spin density in nuclear collisions, demonstrating that it undergoes event-by-event fluctuations arising from the finite number of participating nucleons. The resulting net spin polarizes the fireball and can generate potentially measurable spin correlations in $\Lambda$ hyperon decays. Two essential features of these effects are that they should be largely independent of beam energy and scale inversely with system size. Consequently, the signal is expected to be significantly stronger in O+O than in Pb+Pb collisions, while it should be nearly identical in O+O data from both RHIC and LHC collision energies.

A robust discovery of initial-state spin fluctuations would thus require: (i) that $v_\Lambda$ should reach the percent level, well above any contribution from vorticity; (ii) that this should persist even when the correlated $\Lambda$ hyperons are separated by a large rapidity gap. Establishing such a discovery would open a new experimental program on the initial states of heavy-ion collisions, allowing stringent tests of spin conservation and relaxation, and offering novel methods to probe the spin structure of nuclei at high energy.

%
%

\end{document}